\documentclass[preprint]{aastex}

\usepackage{epsfig}
\usepackage{lscape}
\usepackage{mathptmx} 

\newcommand{\myemail}{s.dzib@crya.unam.mx}

\newcommand{\dechms}[4]{$#1^{\rm h}#2^{\rm m}#3\mbox{$^{\rm s}\mskip-7.6mu.\,$}#4$}

\shorttitle{A Preliminary distance to EC 95}
\shortauthors{Dzib et al.}

\begin{document}

\title{VLBA determination of the distance to nearby star-forming regions\\
IV. A preliminary distance to the proto--Herbig A\lowercase{e}B\lowercase{e} star EC 95 in the Serpens Core}

\author{Sergio Dzib, Laurent Loinard}

\affil{Centro de Radioastronom\'{\i}a y Astrof\'{\i}sica, Universidad
Nacional Aut\'onoma de M\'exico\\ Apartado Postal 3-72, 58090,
Morelia, Michoac\'an, M\'exico (\myemail)}

\author{Amy J.\ Mioduszewski}

\affil{National Radio Astronomy Obserbatory, Domenici Science Operations Center,\\
1003 Lopezville Road, Socorro, NM 87801, USA}

\author{Andrew F.\ Boden}

\affil{Division of Physics, Math, and Astronomy, California Institute of Technology, \\ 
      1200 E California Blvd., Pasadena CA 91125, USA}

\author{Luis F. Rodr\'{\i}guez}

\affil{Centro de Radioastronom\'{\i}a y Astrof\'{\i}sica, Universidad
Nacional Aut\'onoma de M\'exico\\ Apartado Postal 3-72, 58090,
Morelia, Michoac\'an, M\'exico}

\and

\author{Rosa M.\ Torres}

\affil{Argelander Institute for Astronomy, University of Bonn, Auf dem H\"ugel 71, 53121 Bonn, Germany}

\begin{abstract}
Using the Very Long Base Array, we observed the young stellar object EC 95 in 
the Serpens cloud core at eight epochs from December 2007 to December 2009.
Two sources are detected in our field, and are shown to form a tight binary system.
The primary (EC 95a) is a 4--5 $M_\odot$ proto-Herbig AeBe object (arguably
the youngest such object known), whereas the secondary (EC 95b) is most likely 
a low-mass T Tauri star. Interestingly, both sources are non-thermal emitters. 
While T Tauri stars are expected to power a corona because they are convective 
while they go down the Hayashi track, intermediate-mass stars approach the main 
sequence on radiative tracks. Thus, they are not expected to have strong superficial 
magnetic fields, and should not be magnetically active. We review several mechanisms 
that could produce the non-thermal emission of EC 95a, and argue that the observed
properties of EC 95a might be most readily interpreted if it possessed a corona
powered by a rotation-driven convective layer. Using our observations, we show that the trigonometric
parallax of EC 95 is $\pi$ = 2.41 $\pm$ 0.02 mas, corresponding to a distance
of 414.9$^{+4.4}_{-4.3}$ pc. We argue that this implies a distance to the Serpens
core of 415 $\pm$ 5 pc, and a mean distance to the Serpens cloud of 415 $\pm$
25 pc. This value is significantly larger than previous estimates ($d$ $\sim$ 260 pc)
based on measurements of the extinction suffered by stars in the direction of 
Serpens. A possible explanation for this discrepancy is that these previous
observations picked out foreground dust clouds associated with the Aquila Rift 
system rather than Serpens itself.
\end{abstract}


\section{Introduction}
An accurate knowledge of the physical properties of young stellar objects (like their 
mass, age and luminosity) is important to constrain theoretical pre--main sequence 
evolutionary models. The determination of these properties, however, depends 
critically on the availability of accurate distances. Unfortunately, since distances to 
regions of star formation are often uncertain by more than 20 or 30\%, errors on 
the luminosity and age of young stars are typically about 70\%. Significant progress 
has been possible in recent years thanks to Very Long Baseline Interferometry (VLBI) 
observations, particularly with the Very Long Base Array (VLBA --Loinard et al.\ 2005, 
2007, 2008; Torres et al.\ 2007, 2009; Menten et al.\  2007; Xu et al.\ 2006). Owing
to the very accurate astrometry delivered by such instruments, trigonometric 
parallaxes (and therefore distances) can be measured very precisely if multi-epoch 
observations spread over a few years are obtained. 

VLBI instruments are only sensitive to high surface brightness emission, and can only 
detect objects where non-thermal processes are at work. Such non-thermal sources
must, therefore, be identified in the regions of interest before their distance can be
measured using multi-epoch VLBI observations. Fortunately, many young stars are
magnetically active, and do exhibit detectable levels of non-thermal radio 
emission.\footnote{Theoretically, only {\it low-mass} young stars are expected to be 
convective, and to have strong superficial magnetic fields. However, the young 6 
$M_\odot$ B4V star S1 in Ophiuchus (Loinard et al.\ 2008) is known to exhibit non-thermal 
radio emission, although theoretical arguments suggest it should be radiative. We will come back to this point
in Sect.\ 4.2.} This type of emission is typically characterized by strong variability, some 
level of circular polarization, and a negative spectral index. Also, for magnetically active
stars, there is a good correlation between X-ray and non-thermal radio emission (Benz
\& G\"udel 2004) so young stars with detectable levels of non-thermal radio emission 
are associated with bright X-ray sources.

In this work, we will focus on the star-forming region associated with the Serpens molecular 
cloud (Strom  et al.\ 1974; see Eiroa 1992 and Eiroa et al.\ 2008 for two recent reviews). More
specifically, we will concentrate on the SVS~4 region (Strom et al.\ 1976), an infrared
cluster of at least 11 pre-main sequence sources  deeply embedded within the Serpens 
core, and one of the densest young stellar sub-clusters known, with a stellar mass density 
of $\sim\, 10^5\, {\rm M}_\odot$ {\rm pc}$^{-3}$  (Eiroa \& Casali 1989). In the direction of 
SVS 4, Preibisch (1998) detected a bright X-ray source ($L_X$
$\sim$ 4 $\times$10$^{31}$ erg s$^{-1}$; Preibisch 2003a) now known to be 
associated with the infrared object EC 95 (Preibisch 2003a, Eiroa \& Casali 1992).
Using IR spectroscopy, Preibisch (1999) showed that EC 95 is $\sim$ K2 star, and a 
comparison with theoretical tracks indicates that it is a very young ($\sim$ 10$^5$ yr), 
intermediate mass star, presumably a precursor of a $\sim$ 4 {\rm M$_{\odot}$} Herbig 
AeBe star. As a consequence, Preibisch (1999) argued that EC 95 is a {\em proto
Herbig AeBe star}. 

\begin{landscape}
\begin{table*}[!htbp]
\small
\begin{center}
\caption{Measured Source Positions and Flux Densities}
  \begin{tabular}{lccccccccc}\hline\hline
Mean UT date& &Julian Day & $\alpha$(J2000.0) & $\sigma_{\alpha}$ &
$\delta$(J2000.0) & $\sigma_\delta$ & $f_\nu$ $\pm$ $\sigma_{f_\nu}$&$\sigma$&T$_b$\\
(yyyy.mm.dd&hh:mm)& & $18^{\rm h}29^{{\rm m}}$& & $+1^{\circ}12'$& & (mJy)&
(mJy beam$^{-1}$)& ($10^7$ K)\\
\hline
\noindent EC 95a& & & & & & & & &\\
2008.11.29&21:34&2454800.40&57\rlap.{$^{\rm s}$}8918723 &0\rlap.{$^{\rm s}$}00000168&46\rlap.{''}110068&0\rlap.{''}000078&1.24 $\pm$ 0.08&0.08&1.71\\
2009.02.27&15:42&2454890.15&57\rlap.{$^{\rm s}$}8921768 &0\rlap.{$^{\rm s}$}00000049&46\rlap.{''}106950&0\rlap.{''}000017&4.01 $\pm$ 0.15&0.08&5.55\\
2009.12.05&21:12&2455171.38&57\rlap.{$^{\rm s}$}8922183 &0\rlap.{$^{\rm s}$}00000070&46\rlap.{''}095333&0\rlap.{''}000032&2.77 $\pm$ 0.16&0.09&4.37\\
\noindent EC 95b     &     &          &                            &
               &                  &               &   &  &\\
2007.12.22&19:38&2454457.32&57\rlap.{$^{\rm s}$}8909548 &0\rlap.{$^{\rm s}$}00000094&46\rlap.{''}108014&0\rlap.{''}000032&1.79 $\pm$ 0.10&0.05&2.14\\
2008.06.29&07:39&2454646.82&57\rlap.{$^{\rm s}$}8909523 &0\rlap.{$^{\rm s}$}00000370&46\rlap.{''}107340&0\rlap.{''}000145&0.47 $\pm$ 0.16&0.09&0.61\\
2008.09.15&02:33&2454724.61&57\rlap.{$^{\rm s}$}8908014 &0\rlap.{$^{\rm s}$}00000087&46\rlap.{''}105880&0\rlap.{''}000029&4.51$\pm$ 0.23&0.11&4.43\\
2008.11.29&21:34&2454800.40&57\rlap.{$^{\rm s}$}8908963 &0\rlap.{$^{\rm s}$}00000168&46\rlap.{''}104204&0\rlap.{''}000078&1.03 $\pm$ 0.08&0.08&1.45\\
2009.02.27&15:42&2454890.15&57\rlap.{$^{\rm s}$}8911287 &0\rlap.{$^{\rm s}$}00000575&46\rlap.{''}103781&0\rlap.{''}000148&0.41 $\pm$ 0.16&0.08&0.45\\
2009.06.03&09:22&2454985.89&57\rlap.{$^{\rm s}$}8910820 &0\rlap.{$^{\rm s}$}00000408&46\rlap.{''}104144&0\rlap.{''}000096&1.57 $\pm$ 0.24&0.11&1.96\\
2009.08.31&03:35&2455074.65&57\rlap.{$^{\rm s}$}8909107 &0\rlap.{$^{\rm s}$}00000085&46\rlap.{''}103210&0\rlap.{''}000030&3.11 $\pm$ 0.14&0.07&4.19\\
\hline\hline
  \label{tab:pa}
  \end{tabular}
\end{center}
\end{table*}
\end{landscape}

Rodr\'{\i}guez et al.\  (1980) and Eiroa et al.\ (1995) detected EC 95 in VLA observations 
as an unresolved and very strong, radio source (one of the strongest in the survey of
Eiroa et al.\ 2005). This radio emission is strongly variable, and has a negative 
spectral index, suggesting a non-thermal (gyrosynchrotron) origin (Smith 1999). In order 
to constrain further the origin of that radio emission, Forbrich et al. (2007) obtained 
combined high sensitivity Very Large Array (VLA) and High Sensitivity Array (HSA; a 
heterogeneous VLBI instrument that includes the VLBA, the VLA, and the Effelsberg and 
Green Bank single-dish telescopes) observations. The source is detected with the VLA as 
a compact $\sim$ 0.5 mJy source, but not on the long baselines of the HSA (at a 3$\sigma$ 
upper limit of about 50 $\mu$Jy). They argue that the emission detected at the VLA traces 
free-free radiation from an ionized wind that might absorb the emission from the underlying 
active magnetosphere. In the present article, however, we will describe new multi-epoch 
VLBA observations of EC 95, in which compact non-thermal emission {\em is} detected on 
very long baselines. We will use these data to estimate the trigonometric parallax of that source, 
and of the Serpens core as a whole. As discussed by Eiroa et al.\ (2008), the distance to 
Serpens has been a matter of some controversy over the years, with estimates ranging from 
210 {\rm pc} to 700 {\rm pc} (see Sect.\ \ref{sect:distance}).  

\begin{figure}[!t]
\begin{center}
\includegraphics[width=0.4\linewidth,angle=-90]{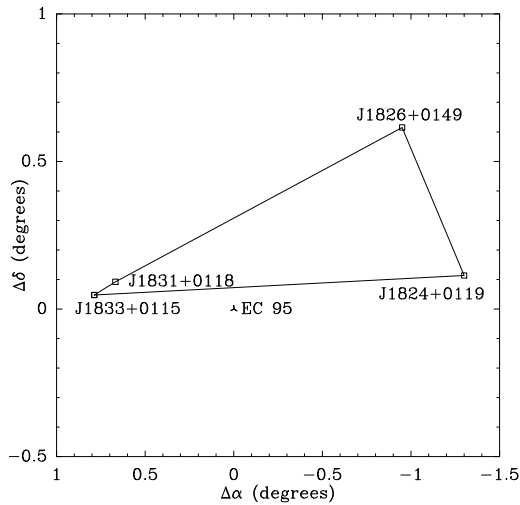}
\end{center}
\caption{Relative position of the astronomical target, as well as of the main 
and secondary calibrators.}
\end{figure}

\section{Observations and data Calibrations}

We present a series of eight radio continuum observations obtained at 3.6 {\rm cm} 
(8.42 {\rm GHz}), using the VLBA of the National Radio Astronomy Observatory (NRAO)
between December 2007 and December 2009. The first observation (in December 2007)
was designed as a detection experiment. Since the source was successfully detected, we 
then initiated a series of multi-epoch observations starting in June 2008. The separation 
between successive observations in those multi-epoch data was about three months 
(Tab.\ 1).

We used the quasar J1833+0115 as main phase calibrator; it is located at 0.79 degrees to 
the target, EC 95. Also, we used the quasars J1832+0118, J1826+0149 and J1824+0124
(located at 0.68, 1.08 and 1.30 degrees from the target, respectively) as secondary phase 
calibrators (Fig.\ 1). For astrometry, it is always desirable for the main calibrator 
to be strong and close to the target. In this case, we used J1833+0115 rather than J1832+0118 
as primary calibrator because the former, although slightly farther from the source, is significantly 
stronger than the latter. Each observation consisted of series of cycles with two minutes spent on 
source, and one minute 
spent on J1833+0115. Every $\sim$ 30 minutes we observed the secondary phase calibrators, 
spending one minute in each. The data were edited and calibrated using the Astronomical Image 
Processing System (AIPS; Greisen 2003). The basic data reduction followed the standard VLBA 
procedure for phase-referenced observations, including the multi-calibrator schemes. It was 
described in detail in Loinard et al. (2007) and Torres et al.\ (2007). 

Even after the careful calibration described above, systematic phase errors, which adversely affect
the astrometric quality of the data, are still present. Above 5 GHz, the largest effects come from 
remaining tropospheric calibration errors due to imperfections in the atmospheric model used by 
the VLBA correlator, and from inaccuracies in the clocks used at each antenna (see Reid \& Brunthaler 
2004).  As a consequence of these errors, the phase correction adequate for the target are slightly different from those
determined using the calibrators. To correct for these effects, one must measure the so-called
group delay (i.e.\ the rate of phase change with frequency). This can be done by observing a dozen
quasars distributed over the entire sky over a wide range of frequencies (Reid \& Brunthaler 2004). From such measurements,
the variable clock delays and the tropospheric term can be obtained, and the systematic phase
errors remaining after standard calibration can be minimized.  We applied this strategy to the last
seven epochs of our data\footnote{Since the first observation was designed as a detection
experiment, it did not include the additional calibration scans required to measure the group delay.}
using observations of multiple ICRF quasars obtained at the beginning, the middle, and the end of 
each observing session. Once the group delays were measured using these data, they were applied
using the dedicated task DELZN in AIPS (see Mioduszewski \& Kogan 2009). 

After the data were calibrated as described above, the visibilities were imaged with a pixel size of 50 
$\mu$as using a weighting scheme intermediate between natural and uniform (ROBUST = 0 in AIPS). 
The rms noise levels in the final images were 0.08 -- 0.10 mJy beam$^{-1}$ (Tab.\ 1).

\begin{figure*}[!ht]
\begin{center}
\includegraphics[width=0.45\linewidth,angle=-90]{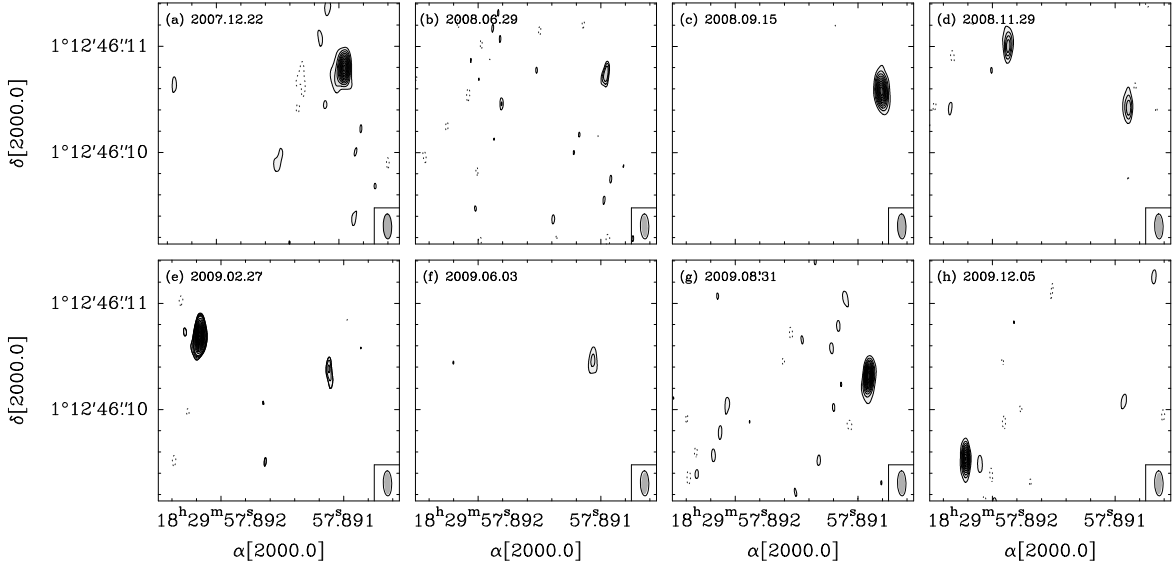}
\end{center}
\caption{Images of EC 95 at all eight epochs. The synthesized beam for each observation
is shown at the bottom-right of each panel. The contour levels are -3, 3,  9, 12, etc.\ times
the noise level in each image (those noise levels are given explicitly in Tab.\ 1). Moreover,
contours at 4 and 5 $\sigma$ were added to panels (b) and (e) to bring out EC 95b,
which was detected only at 5$\sigma$ at those epochs.}
\end{figure*}

\section{Results}

\subsection{Structure and properties of the emission}

In six of our eight observations, a single source was detected in the VLBA images, whereas
two sources were simultaneously detected in the other two (Fig.\ 2). As Fig.\ 2 shows, a source 
is detected at $\alpha~\sim~$\dechms{18}{29}{57}{891} at seven epochs, whereas another 
source is detected at $\alpha~\sim~$\dechms{18}{29}{57}{892} only three times (panels (d), 
(e) and (h) in Fig.\ 2). The location of both sources on the plane of the sky appears to change
with time (Fig.\ 2), so it is clear that neither of them is an extragalactic background object. 
Moreover, a comparison between panels (d) and (e) in Fig.\ 2 clearly shows that both sources 
are moving on the plane of the sky in a similar fashion. Thus, they are most likely associated 
with one another. It would, indeed, be extremely unlikely for two Galactic radio sources separated 
by a mere 15 mas to be unrelated. We conclude that both radio sources detected in our VLBA 
observations are associated with the protostellar object EC 95.  For reasons that will become 
apparent momentarily, we shall call the source at  $\alpha~\sim~$\dechms{18}{29}{57}{891} 
EC 95b, and that at $\alpha~\sim~$\dechms{18}{29}{57}{892} EC 95a. 

Both sources are found to be very variable (Fig.\ 3). While they both can reach a flux of about 
4 mJy, they sometimes remain undetected at levels below about 0.2 mJy. The brightness
temperature of the emission is typically of the order of 10$^7$ K (last column in Tab.\ 1). All 
these properties demonstrate that the emission is non-thermal, and strongly suggest that each 
radio source is associated with a flaring star. Interestingly, no circular polarization was detected 
in our observations --to an upper limit of about 5\%. This is unlike the situation in other magnetically 
active young stars (e.g.\ Loinard et al.\ 2007, 2008, 2009, Torres et al.\ 2007, 2009) and likely 
indicates that the magnetic field topology on the active stars in EC 95 is complex.

\subsection{Astrometry}

The positions of the sources in our VLBA images were determined using a two-dimensional 
Gaussian fitting procedure (task JMFIT in AIPS) and are given in Tab.\ 1. JMFIT provides an 
estimate of the position error based on the expected theoretical astrometric precision of an 
interferometer (Condon 1997); these errors are quoted in columns 4 and 6 of Tab.\ 1. To 
obtain the astrometric parameters from these data, we used the SVD--decomposition fitting 
scheme described by Loinard et al.\ (2007). The necessary barycentric coordinates of the 
Earth, as well as the Julian date of each observation, were calculated using the Multi-year 
Interactive Computer Almanac (MICA) distributed as a CD ROM by the US Naval Observatory.
The reference epoch was taken at the mean of our observations: JD 2454765.98 $\equiv$ 
J2008.90.

Since EC 95a is only detected three times, it would be very hazardous to 
fit the observed positions with a combination of parallax and proper motions.
Thus, we will concentrate here on the EC 95b component. Two fits were 
performed. In the first one, we assumed a linear and uniform proper motion. 
The best fit under this assumption is shown on the left panel of Fig.\ 4, and
yields the following astrometric elements:

\begin{eqnarray*}
\alpha_{J2008.9} & = & 18^{{\rm h}}29^{{\rm m}}57\rlap.{^{\rm s}}890983\pm 0.000009\\
\delta_{J2008.9} & = &\ 1^{\circ}12^{'}46\rlap.{''}1055 \pm 0.0001\\
\mu_\alpha \cos{\delta}&=&0.72 \pm 0.25\ {\rm mas\ yr}^{-1}\\
\mu_\delta&=&-3.61 \pm 0.20\ {\rm mas\ yr}^{-1}\\
\pi&=&2.30 \pm 0.19\ {\rm mas}.
\end{eqnarray*}

The corresponding distance is 435.2$^{+38.8}_{-32.9}$ {\rm pc}. Note, however, that the 
post-fit rms are somewhat large: 0.28 and 0.21 {\rm mas} in right ascension and declination, 
respectively. Indeed, Fig.\ 4 shows that the observed positions often do not coincide with 
the positions expected from the best fit. 

We saw earlier that two sources are detected in our VLBA images, and that those
sources are likely to trace two associated active stars. Under these circumstances, it 
is to be expected that they will be in gravitational interaction, and that their motions will 
be accelerated rather than uniform. As a consequence, we performed another fit 
including a uniform acceleration term (similar to that included in the fit to the T Tau 
data; see Loinard et  al.\ 2007). This fit is shown on the right panel of Fig.\ 4, and 
yields the following parameters:

\begin{eqnarray*}
\alpha_{J2008.9} & = & 18^{{\rm h}}29^{{\rm m}}57\rlap.{^{\rm s}}890966\pm 0.000002\\
\delta_{J2008.9} & = &\ 1^{\circ}12^{'}46\rlap.{''}10532 \pm 0.00008\\
\mu_\alpha \cos{\delta}&=&0.70 \pm 0.02\ {\rm mas\ yr}^{-1}\\
\mu_\delta&=&-3.64 \pm 0.10\ {\rm mas\ yr}^{-1}\\
a_{\alpha}\cos{\delta}&=& 1.95 \pm 0.05\ {\rm mas\ yr}^{-2}\\
a_\delta&=& 1.41 \pm 0.36\ {\rm mas\ yr}^{-2}\\
\pi&=&2.41 \pm0.02\ {\rm mas}
\end{eqnarray*}

This correspond to a distance of 414.9$^{+4.4}_{-4.3}$ pc, consistent with the value 
obtained without using the acceleration terms,  but nearly ten times more accurate.  
The post-fit rms is 0.02 mas in right ascension and 0.10 mas in declination, significantly 
better than those obtained when no acceleration is included (indeed, Fig.\ 4b shows 
that the agreement between the data and the fit is much better when acceleration is 
included). The  reduced $\chi^2$ obtained in right ascension using
the errors delivered by JMFIT is 0.6, suggesting that no significant systematic
errors remain in our data along that axis. In declination, however, a systematic contribution
of  0.127 mas has to be added quadratically to the errors given by JMFIT to obtain
a reduced $\chi^2$ of 1. All the errors quoted here include this systematic contribution.

\begin{figure*}[!ht]
\begin{center}
\includegraphics[width=0.35\linewidth,angle=-90]{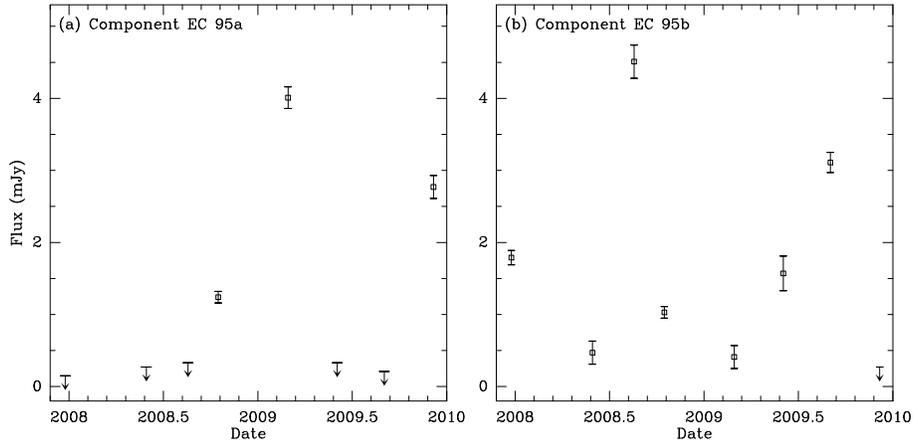}
\end{center}
\caption{Time evolution of the 3.6 cm flux of EC 95a and EC 95b in our VLBA data. 
The upper limits are at 3$\sigma$.}
\end{figure*}

\subsection{Binarity}

Adding free parameters to a fit will always improve the agreement between 
the fit and the data. Therefore, the improvement in our fit when acceleration 
terms are included does not necessarily guarantee that an accelerated
trajectory provides a better description of the data. The premise to include 
an acceleration term was the existence of a gravitational interaction between
the two sources in our field. Accordingly, we must now verify that the acceleration 
vector obtained from the fit does point toward the second radio source.

To examine this issue, we subtracted the parallactic contribution from the observed
positions of both EC 95a and EC 95b (Fig.\ 5a). While the trajectory of EC 95b 
is significantly curved, that of EC 95a appears to be linear and uniform to a very 
good degree of precision. This confirms that EC 95a and EC 95b are at the same distance, 
since it would require an extraordinary coincidence to obtain a uniform motion
for EC 95a after applying the parallax correction of EC 95b if the two sources were
not at the same distance. Moreover, the 
acceleration vector (of EC 95b) found in our fit appears to point almost exactly 
towards the expected position of  EC 95a at the median epoch of our observations
(Fig.\ 5a). This is exactly what would be expected if EC 95a and EC 95b formed a gravitationally 
bound binary system. Finally, since the trajectory of EC 95b is curved and accelerated,
whereas that of EC 95a is linear and uniform, EC 95a must be significantly
more massive than EC 95b. We conclude that EC 95a and EC 95b do constitute
a binary system, where EC 95a is the primary and EC 95b the secondary. This is, indeed, 
the reason why we ascribed those names to the sources in the first place.

To further characterize the system, it is useful to calculate the separation
between EC 95 a and EC 95b as a function of time. Since the two sources
are detected simultaneously at only two epochs, this can be unambiguously done 
only at those two epochs. However, since the motion of EC 95a appears to be
very nearly linear and uniform, we can estimate the position of the primary at the
other five epochs when the secondary is detected. Those positions are shown as
blue solid circles in Fig.\ 5a. For epochs 6 and 7, the estimation of the position of
the primary only involves an interpolation. However, for the first three epochs, 
somewhat more uncertain extrapolations are needed. Using these estimates for
the position of the primary, we calculated the separation between the two members
of the system shown in Fig.\ 5b. It is clear from that figure that our observations
only cover a fairly small fraction of the orbit, and that any orbit modeling will be
very uncertain. A poorly constrained, but plausible fit (see below) is shown as a 
solid curve. The dotted curve corresponds to the parabolic path predicted by our 
uniformly accelerated fit. The fact that those two trajectories are nearly indistinguishable 
over the course of our observations justifies a posteriori the use of a uniformly
accelerated motion in our astrometric fit. 

A strict lower limit to the mass of the primary can be found from the magnitude of the
acceleration vector determined earlier. If $m$ and $M$ are the masses of the secondary 
and the primary, respectively, Newton's law applied to the secondary yields:

\begin{equation}
m a = {G m M \over r^2} \Rightarrow M = {a r^2 \over G}
\end{equation}

\noindent
where $a$ is the magnitude of the acceleration (of the secondary), and $r$ is the true 
separation between the sources. Of course, we measure only quantities projected onto 
the plane of the sky, so the measured separation and acceleration are only lower limits.
From the measured values ($a_{min}$ = 2.4 mas yr$^{-2}$ $\equiv$ 0.015 cm s$^{-2}$
and $r_{min}$ = 15.8 mas $\equiv$ 9.86 $\times$ 10$^{13}$ cm),
we obtain a minimum mass for EC 95a of 1.1 $M_\odot$. The mass goes roughly as
$\cos^3{i}$, so if the inclination were 45$^\circ$, the mass of EC 95a would be about
3 $M_\odot$.

To obtain an alternative mass estimate, we model the component relative
astrometry data (Tab.\ 1; Fig.\ 5b) with a Keplerian orbit.  As these
data apparently cover only a small ($\sim$ 10\%) fraction of the EC 95 orbit,
we constrained the orbit model to be circular ($e$ = 0); this assumption
is not physically motivated, but the data do not presently support a
more complex interpretation.  The result of our modeling is shown as a
solid line in Fig. 5b. The semi-major axis of this preliminary orbit is
31 $\pm$ 9 mas, its inclination is $i$ = 60$^\circ$ $\pm$ 10$^\circ$, 
and the orbital period is 16.5 $\pm$ 5.0 yr.  These orbit parameters imply 
a total mass for the system of $M$ = 8$^{+27}_{-6}$ $M_\odot$. The large 
error on the upper limit reflects the fact that the mass goes roughly as 
$\cos^3{i}$, and increases rapidly for $i$ above 45$^\circ$. Even with this 
consideration, however, the orbit parameter and mass errors are likely to 
be underestimated because of the non-physical assumption of a circular 
orbit required in this very preliminary modeling.  

\begin{figure*}[!ht]
\begin{center}
\includegraphics[width=0.35\linewidth,angle=-90]{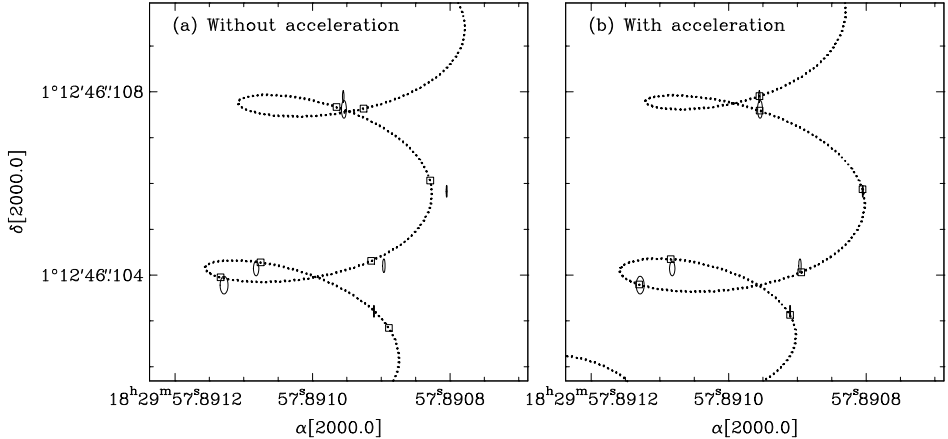}
\end{center}
\caption{Observed positions of EC 95b and best astrometric fits (a) without
acceleration terms, and (b) with acceleration terms. The squares indicate the
position of the source expected from the fits at each epoch, and the ellipses
show the actual observed positions. The size of the ellipse represents the error
on the measurement.}
\end{figure*}

\section{Properties of EC 95 and SVS 4}

\subsection{Nature of EC 95}

From near infrared spectroscopy, Preibisch (1999) estimated a spectral
type K2 $\pm$ 2 for EC 95, and an extinction-corrected bolometric luminosity 
$L$ = 60$^{+30}_{-20}$ $L_\odot$. This was obtained assuming a distance
to Serpens of 310 pc; using the value of the distance obtained here increases
the luminosity to 108$^{+54}_{-36}$ $L_\odot$. More recently, Doppmann et 
al.\ (2005) estimated the effective temperature of EC 95 to be 4400$^{+115}_{-57}$ 
K, in agreement with the spectral type proposed by Preibisch (1999), and its
luminosity (calculated assuming a distance of 259 pc) to be 23 $L_\odot$. 
Scaled to the distance determined here, this would imply a luminosity of
59 $L_\odot$, also consistent with the determination of Preibisch (1999).
The extinction determined by Preibisch (1999) is $A_V$ = 36 $\pm$ 2; an
alternative determination by Pontoppidan et al.\ (2004) yields $A_V$ $\sim$ 
37. 

The mass and the age of EC 95 can be estimated by placing it on an HR 
diagram. Using that method, Preibisch (1999) estimated a mass of about 
4 $M_\odot$ and an age of 0.2$^{+0.2}_{-0.1}$ Myr. Pontoppidan et al.\ 
(2004) proposed a similar mass (3.5 $M_\odot$), and a marginally larger 
age (0.4 Myr) because the shorter distance they used implies a correspondingly 
smaller luminosity. In any case, all these characteristics suggest that EC 95 
is a very young precursor of a $\sim$ 4 $M_\odot$ star. Using our new distance 
determination, the position of EC 95 on the HR diagram moves up somewhat, and 
its position suggests a mass of about 5 $M_\odot$, and an age of about 0.2 Myr, 
confirming that EC 95 is a precursor of an intermediate mass star. In spite of their 
large uncertainties, the dynamical mass estimates given above are also consistent
with that suggestion. Our observations further show that EC 95 is a binary system 
where the primary is significantly more massive than the secondary. As a 
consequence, we identify the proto-Herbig AeBe star with EC 95a, and argue that 
EC 95b must be a low-mass T Tauri companion. It is clear that in such a situation, 
both the bolometric luminosity and the spectral type of the system will be almost 
entirely dominated by the more massive component. We should point out, also, that 
another young star (EC 92), located about 5$''$ ($\sim$ 2000 AU) to the north 
of EC 95, has been argued to be gravitationally bound to it (e.g.\ Haisch et al.\ 2002). 
If that were indeed the case, then EC 92/EC 95 would constitute a rare example of 
a very young, intermediate-mass, hierarchical triple system.

In spite of its youth, EC 95 does not appear to contain large quantities of
circumstellar material. It shows only a modest mid-infrared excess (Preibish 
1999, Haisch et al.\ 2002, Pontoppidan et al.\ 2004), little veiling (Doppmann 
et al.\ 2005), and fairly weak CO overtone rovibrational absorption lines. 
This has led some authors to classify it as a Class II source (e.g.\ Eiroa et al.\
2005), although others have favored a flat spectrum (e.g.\ Harvey et al.\ 2007)
or even Class 0/I classification (Winston et al.\ 2007). We note, in particular,
that the fairly large rotational velocity of EC 95 (56 km s$^{-1}$; Doppmann 
et al.\ 2005) would be more consistent with a Class I or flat spectrum classification.
It is interesting to compare these characteristics with those of the nearby, 
possibly associated, source EC 92. From its location on the HR diagram, 
Preibisch (1999) estimate that EC 92 is a 0.5 $M_\odot$ star with an age of 
the order of 10$^5$ yr. This suggests that EC 92 and EC 95 are nearly 
coeval (as would be natural if they belonged to a common multiple system).
Unlike EC 95, however, EC 92 does exhibit a significant mid-infrared excess,
and has been almost unanimously classified as a flat-spectrum Class I source 
(see e.g.\ Pontoppidan et al.\ 2004). Thus, 
while EC 92 and EC 95 appear to have similar ages and might be physically
associated, they differ significantly in their circumstellar content. The fact that 
EC 95 is a tight binary system could naturally explain the relative paucity of its 
circumstellar content because tidal forces tend to be effective at clearing out 
circumstellar material. In particular, any disk existing around the members of 
the EC 95 system are expected to be truncated down to a radius at most about 
a third of the physical separation between the stars: $\sim$ 5 mas ($\equiv$ 2 AU). 

\subsection{Origin of the non--thermal emission in EC 95}

Low-mass young stars (i.e.\ T Tauri stars) have long been known to often
be magnetically active (e.g.\ Feigelson \& Montmerle 1999). The accepted
explanation for this activity is based on a scaled-up version of the situation
with the Sun. While they move down the Hayashi track towards the main
sequence, stars less massive than about 2 $M_\odot$ are fully convective
(e.g.\ Palla \& Stahler 1993). As a consequence, they can generate strong 
superficial magnetic fields ($\sim$ 1 kG) through the dynamo mechanism
(Parker 1955). This leads to the appearance of magnetic loops anchored 
on the stellar surface, and extending up to a maximum height of a few stellar 
radii. When two loops interact, flares associated with magnetic reconnection 
events can occur, leading to the sudden release of large quantities of energy 
initially stored in the magnetic field. Part of that energy can accelerate electrons 
initially trapped in the loops to mildly relativistic speeds. These electrons then 
spiral in the strong ambient magnetic fields, and generate gyrosynchrotron 
radiation that can be detected at radio wavelengths. The energy released during 
reconnection also serves to heat the stellar corona to temperatures sufficient
to generate detectable thermal brehmsstrahlung X-ray emission. 

Stars more massive than about 2--3 $M_{\odot}$ are expected to move towards 
the main sequence on radiative tracks, and are not anticipated to have strong
superficial fields. Such stars are, therefore, not expected to be magnetically
active. X-ray emission as well as non-thermal radio emission have been 
detected from massive O and WR stars (e.g.\  Bergh\"ofer et al. 1997, Pittard \& 
Dougherty 2006), but are believed to be the result of shocks in winds and 
wind interactions. For intermediate mass stars (spectral type A and late B), 
however, the situation appears somewhat more uncertain. Those stars should
not be magnetically active, and have no strong winds. Yet  a small fraction 
of them (perhaps about 5\%; Montmerle et al.\ 2005) do show evidence of
strong magnetic activity (Stelzer et al.\ 2005, 2006; Wade et al.\ 2009; 
Hubrig et al.\ 2009). Several explanations have been put forward, but
arguably the most plausible one is that the magnetic activity is in fact 
associated with a low-mass companion rather than with the intermediate 
mass primary itself (e.g.\ Stelzer et al.\ 2005, 2006).  Several recent 
X-ray observations, however, might be more easily interpreted if the 
Herbig AeBe stars themselves were magnetically active (e.g.\ Telleschi 
et al\ 2007, G\"unther \& Schmitt 2009, Huenemoerder et al.\ 2009).

Until the present observations, EC 95 was believed to be a single 
intermediate-mass young star, and was, therefore, not expected to show
magnetic activity.  Indeed, Smith et al.\ (1999) and Giardino et al.\ (2007)
did proposed that the non-thermal radio emission as well as the X-ray 
emission from that source might be provided by 
a low-mass companion rather than by the AeBe protostar. Preibisch 
(1999) also mentioned that a low-mass companion could be at the origin
of the X-ray emission, though he initially did not consider this possibility 
particularly likely since the X-ray luminosity of EC 95 based on ROSAT
observations appeared to be nearly three orders of magnitude larger
than that of typical T Tauri stars. However, more recent XMM-Newton 
observations have shown that the X-ray luminosity of EC 95 is significantly
smaller than originally thought, and is in fact within the range of normal 
T Tauri stars (albeit near the top of the X-ray luminosity function). Thus, 
it appears plausible again that the X-ray emission might come from a 
low-mass companion.

\begin{figure*}[!ht]
\begin{center}
\includegraphics[width=0.4\linewidth,angle=-90]{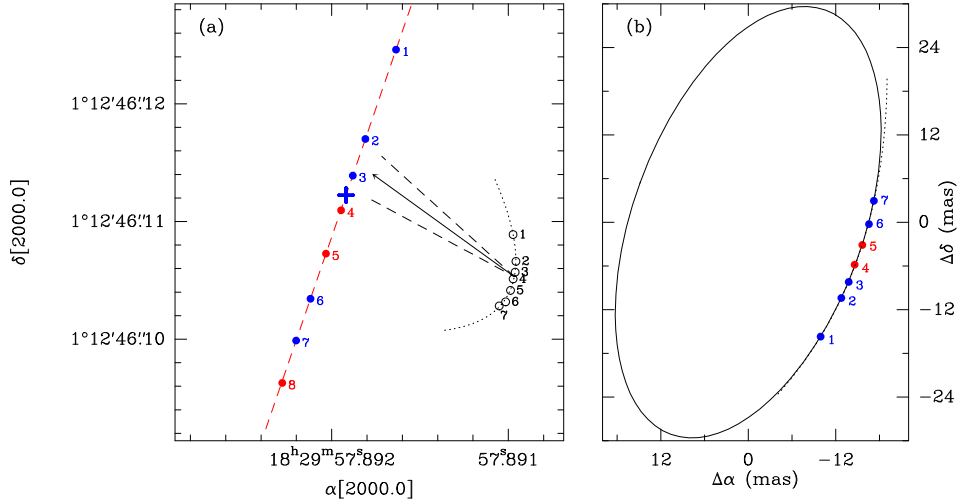}
\end{center}
\caption{(a) Positions of EC 95a and EC 95b as a function of time
after removing the parallactic component. The black empty circles 
show the positions of EC 95b and the filled circles correspond to 
 EC 95a. The epoch is indicated by a number on the side of each 
 symbol. For EC 95a, the epochs for which the positions are measured 
 are shown in red, whereas the blue symbols are for the epochs where
the position was obtained using the (uniform) fit shown as a dashed 
line. (b) Relative position between EC 95b and EC 95a, deduced from 
the absolute positions shown in panel (a). The ellipse shows the best 
fit to the data points assuming an intrinsically circular orbit (see text).
In both panels, the dotted curve shows the best uniformly accelerated 
fit to the original data.}
\end{figure*}

Our observations have revealed that EC 95 is a binary system most
likely composed of a $\sim$ 4-5 $M_\odot$ primary, and a low-mass
T Tauri companion. Moreover, the low-mass companion (EC 95b) 
appears to be usually brighter at radio wavelengths than the more 
massive primary (Tab.\ 1; Fig.\ 3). Given the existing correlation 
between the X-ray and the radio emission of T Tauri stars, one would 
expect EC 95b to also be the brightest member of the system at X-ray 
wavelengths. Thus, the magnetic activity from EC 95 is likely to be largely 
dominated by a low-mass companion, rather than by the intermediate 
mass primary, as proposed by Smith et al.\ (1999). However, our observations 
show that the intermediate mass star EC 95a, although on average somewhat 
weaker than EC 95b, is {\it also} a non-thermal radio emitter.\footnote{One could 
argue that the magnetic activity might come from yet another low-mass companion
of EC 95a, on a very tight orbit. We consider this possibility highly unlikely because its presence would 
generate a wobble in the measured positions of EC 95a and the relative position 
of EC 95b. We find no such wobbling at the level of about 0.2 mas, so the semi-major
axis of the companion would have to be less than about 0.1 AU, or 20 
$R_\odot$.} Thus, the
existence of a low-mass companion does not entirely solve the 
mystery of the magnetic activity in the intermediate mass system
EC 95. A mechanism capable of generating magnetic activity on the
4--5 $M_\odot$ star EC 95a still has to be identified. 

Two classes of processes have been proposed to explain intrinsic 
magnetic activity in intermediate mass stars. The first one invokes 
wind or accretion shocks 
(see Skinner \& Yamaushi 1996 for a detailed discussion). For 
reasonable values of the wind velocity and of the free-fall speed 
($\sim$ 500 km s$^{-1}$), these processes generate plasma
temperatures ($kT$) below 1 keV. If a single temperature is assumed,
the best fit to the X-ray spectrum of EC 95 is obtained for $kT$ = 2.8 keV 
(Preibisch 2003a), a value too high to be reconciled with the idea of
a shock origin. However, a slightly better fit to the spectrum is obtained 
for a two-temperature plasma model with $kT_1$ = 0.7 keV, and $kT_2$ = 
3.0 keV. Could the high temperature plasma trace the T Tauri star, and 
the low-temperature component a shock associated with the intermediate 
mass star? In this scenario, the variability of the radio emission from
EC 95a would be ascribed to changes in the accretion/ejection activity
of the star, so fairly large variations in that activity would be required. 
During episodes of enhanced activity, the X-ray and radio
emissions would be expected to rise, so the relative contribution 
of the low-temperature component would be expected to increase. Thus
a comparison between X-ray observations obtained during a low- and
a high-state of EC 95a would provide a valuable test of that possibility.

A second mechanism capable of producing magnetic activity in
an intermediate-mass star is related to the existence of a thin temporary 
convective layer on the surface of the star. Two mechanisms that might lead 
to the appearance of such a layer are substellar deuterium burning during the 
very early stages of stellar evolution (Palla \& Stahler 1993), and differential rotation 
in a rapidly rotating star (Tout \& Pringle 1995, Skinner et al.\ 2004). 
For a 3--4 $M_\odot$ star, Palla \& Stahler (1993) estimate that deuterium 
burning could last for about 10$^5$ yr. A corona powered by rotation, 
on the other hand, could survive for up to about 10$^6$ yr. Since the age 
of EC 95 is of the order of 10$^5$ yr, both mechanisms could in principle 
maintain a corona over the required timescale, although a rotation-driven
corona would accommodate the age of EC 95a somewhat more comfortably. 
Independently of the mechanism creating it, a thin convective layer would 
power a magnetic activity similar to that occurring in T Tauri stars. Thus, the 
flare-like variability observed in our radio observations of EC 95a (Fig.\ 3a) 
would be naturally interpreted in this scenario. We note, finally,  that EC 95a 
is a fairly fast rotator ($V \sin i$ = 56 $\pm$ 1 km s$^{-1}$; Doppmann et al.\ 
2005) as would be required for a temporary convective layer to be powered 
by differential rotation. We conclude that all the known properties of EC 95a
would be readily interpreted if its magnetic activity were due to a rotation-driven 
convective layer, and we favor this mechanism over alternative ones. 

\subsection{Luminosities of EC 95}

In the previous section, we have argued that, although EC 95a is also
magnetically active, the radio and X-ray emissions from EC 95 are 
dominated by the contribution of EC 95b, which we identify with a T 
Tauri star. We now examine whether or not the relation between the 
X-ray, the radio, and the bolometric luminosities of EC 95b are consistent 
with that possibility.

Recent surveys (e.g.\ G\"udel et al.\ 2007) have shown that the X-ray 
luminosity of T Tauri stars can vary from about 10$^{28}$ to about
10$^{31}$  erg s$^{-1}$. Moreover, X-ray luminosities in excess of 
10$^{32}$ erg s$^{-1}$ have been observed during flares (e.g.\ Preibisch 
2003b). Scaled to the distance found here, the X-ray luminosity\footnote{Here 
we scale the value of the 
luminosity obtained by Preibisch (2003a) using a single-temperature fit. A 
two-temperature model yields a slightly better fit, and a somewhat higher 
luminosity. However, the temperature of the second component in that fit 
($kT$ $\sim$ 0.7 keV) is quite atypical of T Tauri stars. If that second 
component is real, it is likely associated with a different kind of structure, 
such as an accretion shock (see Sect.\ 4.2).} of EC 95 appears to be about 
7 $\times$ 10$^{31}$ erg s$^{-1}$. This, of course, includes both sources 
in the system. If our interpretation is correct, however, most of that luminosity
is to be ascribed to EC 95b, which would appear to be somewhat brighter
in X-rays than typical quiescent T Tauri stars. Its X-ray luminosity would be 
more comparable to that of T Tauri stars during flares. 

The typical X-ray to bolometric luminosity ratio for T Tauri stars 
is 5 $\times$ 10$^{-4}$, but it can be as high as a few times 10$^{-3}$
(e.g.\ Preibisch 2003a). Using the {\it total } bolometric luminosity of EC 95, 
we obtain $L_X/L_{bol}$ $\sim$ 1.6 $\times$ 10$^{-4}$, near the low-end
of typical values found in T Tauri stars. However, if the X-ray luminosity 
is dominated by the low-mass T Tauri star EC 95b, one should calculate
the ratio using the bolometric luminosity of EC 95b only. Since the total
bolometric luminosity of EC 95 is largely dominated by the intermediate 
mass primary, the ratio estimated above must be severely underestimated.
For instance, if EC 95b were a 5 $L_\odot$ T Tauri star, then $L_X/L_{bol}$ 
for that star would be 3.5 $\times$ 10$^{-3}$, near the {\it upper} end of the 
range for T Tauri stars.

The mean 3.6 cm flux of EC 95b in the first seven epochs of our observations
is 1.8 mJy ($\equiv$ 1.8 $\times$ 10$^{-26}$ erg s$^{-1}$ cm$^{-2}$ Hz$^{-1}$).
This corresponds to a radio luminosity $L_R$  = 3.8 $\times$ 10$^{17}$
erg s$^{-1}$ Hz$^{-1}$. Thus, the X-ray to radio luminosity ratio for EC 95b 
is $L_X/L_R$ $\sim$ 1.84 $\times$ 10$^{14}$ Hz ($\equiv$ 10$^{14.3}$ Hz). 
This ratio for stellar coronae is typically $L_X/L_R$ $\sim$ 10$^{15.5}$ Hz 
with a dispersion of about 1 dex (G\"udel 2004; Benz \& G\"udel 1994). Thus,
in terms of its radio to X-ray luminosity ratio, EC 95b appears to be within the 
normal range for stellar coronae. 

In summary, if EC 95b is indeed at the origin of most of the magnetic activity
in the EC 95 system, then it would have to be interpreted as a fairly X-ray
bright T Tauri star (with an X-ray luminosity more typical of a flaring than
of a quiescent star) but with otherwise normal global properties (in terms of
$L_X/L_{bol}$ or $L_X/L_R$). We have shown earlier that EC 95 is
a tight binary system. Close binarity is known to enhance magnetic 
activity in stars --for instance, through tidal synchronism, which tends to increase
the angular velocity of the members of binary systems. Thus, it is plausible
that EC 95b is (because of its binarity) a particularly magnetically active,
but otherwise normal, T Tauri star. 

\subsection{A solution to the extinction mystery of EC 95?}

There is an unsolved mystery regarding the extinction to EC 95. 
While the reddening deduced from near infrared photometric observations seems 
to imply a very large absorption ($A_V$ = 36; Preibisch 1999), the extinction obtained 
from the X-ray spectrum is significantly smaller --albeit still very large ($A_V$ $\sim$ 17). 
Several mechanisms have been proposed to resolve this conundrum (see the discussion 
by Preibisch 2003a) but no fully satisfactory explanation has yet been provided. 
The fact that two sources are detected in EC 95 might naturally explain out
this mystery. The infrared luminosity is
clearly largely dominated by the more massive star EC 95a. Thus, the
extinction deduced from infrared observations ($A_V$ $\sim$ 36) is the 
extinction to that specific star. As  discussed earlier, the X-ray emission, 
on the other hand, is likely to be dominated by the low-mass companion 
EC 95b, so the extinction deduced from X-ray data ($A_V$ $\sim$ 17)
refers to that second star (see, indeed, Giardino et al.\ 2007). The different 
values obtained from infrared and X-ray observations, therefore, suggest 
that the extinction to EC 95a is about 20 magnitudes larger than that to 
EC 95b. Since the two sources are separated by only about 15 mas 
($\equiv$ 7 AU), the structure responsible for the excess of emission 
towards EC 95a must be very compact --for instance in the form of a 
dense circumstellar disk. 

Interestingly, such a structure is known to exist in at least one other Herbig AeBe 
candidate: T Tauri Sa. Based on observations of rovibrational lines of CO, Duch\^ene 
et al.\ (2005) have shown that T Tau Sa is surrounded by a compact structure which 
would generate an extinction $A_V$ of about 90 magnitudes if the dust-to-gas ratio 
in that structure were similar to that of the interstellar medium. They interpret that structure as
a nearly edge-on, compact ($R$ $\sim$ 2--3 $R_\odot$) circumstellar disk. 
It is noteworthy that EC 95a and T Tau Sa are similar in several respects. Both are 
most likely young Herbig AeBe stars orbited by a low-mass companion (T Tau Sb, 
and EC 95b, respectively) located at about 10 AU. Thus, in both cases, the circumstellar
disks are expected to be tidally truncated to about 2--3 AU. The rovibrational CO lines in 
EC 95 are somewhat fainter than those in T Tau Sa, and the K-band veiling of EC 95
($r_K$ $\sim$ 0.1; Doppmann et al.\ 2005) is also smaller than that of T Tau Sa ($r_K$
$\sim$ 2; Duch\^ene et al.\ 2003).  Thus, the amount of circumstellar material around
EC 95a must be significantly smaller than that around T Tau Sa. Still, a scaled-down 
version of the compact disk surrounded T Tau Sa, consistent with the fainter CO lines
and smaller veiling of EC 95, might well provide the 20 magnitudes of extinction
required to reconcile the infrared and X-ray extinction determinations.

\subsection{Stellar density of SVS 4}

We mentioned in the introduction that SVS 4 is believed to be one of the
densest sub-clusters in the Milky Way. Given the determination of the distance
to SVS 4 given here, it is useful to re-examine this issue. Eiroa \& Casali (1989)
identified 11 young stellar objects in an angular diameter of $\sim$ 50''. Assuming
a typical mass of 1 M$_\odot$ for each object, they obtained stellar mass densities 
between $4\times10^3$ and $10^{5}$ M$_\odot$ pc$^{-3}$ (the broad range 
reflected the uncertainty on the distance to Serpens). The 11 objects in SVS 4
are now known to be embedded in a common faint nebulosity covering an 
area of $\sim$ 35$''~\times$35$''$ (Eiroa et al. 2008). This implies a diameter of
$\sim$ 0.07 pc for the SVS 4 group. The mass of EC 95 is about 5 $M_\odot$, so
if we assume the other 10 objects in SVS 4 to be 1 $M_\odot$ stars, we obtain
a total mass of 15 $M_\odot$ for the sub-cluster. This yields a stellar mass density 
of about 8$\times$10$^4$ M$_\odot$ pc$^{-3}$. This is consistent with the
estimates of Eiroa \& Casali (1989), and confirms that SVS 4 is one of the
densest concentration of pre main sequence stars known in the Galaxy.

\section{The distance to the serpens core and cloud}\label{sect:distance}

As we mentioned earlier, the distance to Serpens has been a matter
of controversy during the past several decades. In Tab.\ 2, we compile 
the distance estimates obtained by different authors over the years. All these 
estimates are indirect, and most of them rely on spectroscopic parallaxes 
either to estimate the distance directly to Serpens members, or to measure 
extinction as a function of distance for objects in the direction of Serpens. Interestingly, several 
of those studies are based on the same samples of stars (or at least on strongly 
overlapping samples), but reach widely different conclusions. This is the consequence of two 
main sources of uncertainty. The first one is the assignment of a spectral 
type to the stars considered, which sets their absolute magnitudes. The 
second is the determination of the extinction factor $R_V$ which is 
required to translate reddening into extinction. There has been some
discussion about the appropriate value of $R_V$ for Serpens (e.g.\
de Lara et al.\ 1991), and the different assumptions made by different
authors are at least partly responsible for the wide range of distances
reported in the literature. 

In a recent review, Eiroa et al.\ (2008) argue that the most recent 
distance estimates seem to converge towards a value of 230 $\pm$ 
20 {\rm pc} for the Serpens molecular cloud. The case for such a
convergence, however, is not immediately apparent from Tab.\ 2.
In fact, that claim appears to be largely based on the results 
presented by Strai{\v z}ys et al.\ (2003). Yet, these authors do 
not calculate the distance to the Serpens cloud, but the distance to 
the Aquila Rift, which is a large system of dark clouds, that includes 
Serpens but is significantly more extended.\footnote{In total, the Aquila 
Rift covers about 400 square degrees of the sky, whereas Serpens is 
only about 6 square degrees.} Moreover, Strai{\v z}ys et al.\ (2003) only 
claim to have measured the distance to {\it the front edge} of the Aquila 
Rift, which they place at 225 $\pm$ 55 pc. They further argue that the 
depth of the complex might be about 80 pc, so depending on the position 
of Serpens within the complex, its distance might be somewhat larger. 
Thus, from the results of Strai{\v z}ys et al.\ (2003), 230 pc appears
as a lower limit to the distance to Serpens rather than as the most probable
value. In a previous article, Strai{\v z}ys et al.\ (1996) did estimate the 
distance specifically to the Serpens  cloud; they obtained 259 $\pm$ 
37 pc. Their more recent Aquila Rift paper includes 80 of the stars used 
in that 1996 article, as well as other 400 stars distributed over the rest of 
Aquila Rift region. Thus, it does not provide (nor claim to provide) a new 
and independent estimate of the distance to Serpens, but rather --as 
mentioned above-- a measurement of the distance to the front edge of 
the Aquila Rift.

\begin{table*}[htb]
\small
\begin{center}
\caption{Measured distances to the Serpens Molecular Cloud}
\begin{tabular}{lclccl}\hline\hline
Authors &Year & \multicolumn{1}{c}{Method}&Distance&$\sigma_d$ &\multicolumn{1}{c}{Sub-region} \\
     &    &      &  (pc)  &   (pc)    &\\
\hline
Racine&1968&Distance modulus to HD170634  (Optical)    &  440    &   &S68 \\
Strom et al.&1974&Distance modulus to HD170634  (Infrared)   &440&   &S68    \\
Chavarria-K et al.& 1988&Spectroscopy, $uvby\beta$  and JHKLM photometry to four stars &245&30&Serpens Cloud\\
Zhang et al. & 1988&Distance modulus to three stars  &650&180& Serpens Core\\
            &     &Kinematic distance of CO and NH$_3$ lines &600&   & \\
de Lara \& Chavarria& 1989&Infrared photometry to seven stars        &296&34&Serpens Cloud\\
de Lara et al.& 1991&Extinction measurements  of five stars     &311&38&Serpens Cloud\\
Chiar         &1996 &Infrared photometry to seven stars    &425&45&Serpens Cloud\\
Strai{\v z}ys et al.&1996&Vilnius photometry of  105 stars &259&37&Serpens Cloud\\
\hline\hline
   \label{tab:di}
   \end{tabular}
 \end{center}
\end{table*}

The distance obtained here is based on a trigonometric parallax measurement,
and, therefore, does not rely on any assumption regarding the nature of the star, 
or the properties of the interstellar medium in the Serpens and Aquila Rift clouds. There is no question that 
EC 95 is located at 414.9$^{+4.4}_{-4.3}$ pc. On the other hand, we only measured the distance 
to that {\it one star}, whereas previous works based on indirect determinations 
(such as those of Strai{\v z}ys et al.\ 1996) relied on much larger samples. One
must, therefore, address the issue of the relevance of the distance measured here
for the entire Serpens complex. In this respect, it is important to point out that EC 
95 is not just any star located in the direction of Serpens. As we have seen earlier, 
it is a very young Herbig AeBe star. Such intermediate-mass stars are not born in isolation, 
but as part of small clusters. Since EC 95 is one of at least 11 pre-main sequence 
stars in the compact SVS 4 region, there is no doubt that it is physically associated 
with that small cluster (see also Eiroa \& Casali 1989), and that the distance measured 
here is also the distance to SVS 4. In addition, SVS 4 is universally acknowledged to be
part of the Serpens core region (e.g.\ Eiroa \& Casali 1992), and its members
(including EC 95) have always been considered part of the Serpens young
stellar cluster (Eiroa \& Casali 1992). Thus, we argue that the distance to EC 95
measured here also provides an accurate estimate of the distance to the Serpens
core. We note further that the Serpens core is only about 5$'$ across on the plane
of the sky (0.6 pc at a distance of 415 pc). Thus, depth effects are not expected to 
add significantly to the error budget on the determination of the distance to the core.
As a consequence, we argue that the distance to the Serpens core should henceforth
be assumed to be 415 $\pm$ 5 pc.

With an angular  diameter of about 3$^\circ$, the Serpens cloud is much larger than 
the Serpens core. At a distance of 415 pc, this would correspond to a physical 
diameter of about 20 pc. The depth of the Serpens complex is also likely
to be of that order, and depending on the location of the core relative to the rest of
the complex, the mean distance appropriate for the Serpens cloud might be up
to 20 pc less or 20 pc more than the distance to the core. Thus, we argue that a 
conservative estimate of the mean distance to the Serpens cloud would be 
415 $\pm$ 25 pc.

This value is significantly larger than the figure reported by Strai{\v z}ys et al.\ (1996) 
for the Serpens molecular cloud ($d$ = 259 $\pm$ 37 pc). Since the total extent of the 
Serpens cloud on the plane of the sky is only about  20 pc, it is extremely unlikely 
to be 150 pc deep. Instead, we argue that several unrelated regions are likely to 
coexist along the line of sight. As mentioned above, Serpens is one of the many 
dust clouds forming the large structure known as the Aquila Rift. Strai{\v z}ys et
al.\  (2003) have shown that the front edge of that structure is at about 230 pc, and
that its depth is about 80 pc. The discrepancy between our result and that of
of Strai{\v z}ys et al.\ (1996) could be readily explained if the Serpens molecular 
cloud were not physically associated with the Aquila Rift, but instead located behind 
it. In this situation, since the method used by Strai{\v z}ys et al.\ (1996) is sensitive to 
the position of the first obscuring material located along the line of sight, it would 
have naturally picked out the foreground Aquila Rift clouds, rather that the more 
distance Serpens region. Indeed, in their concluding remarks, Strai{\v z}ys et al.\ 
(2003) state that {\it both W40 and the Serpens molecular cloud are seen projected 
on the very dark foreground created by the dust lanes of the Aquila Rift}. Such a 
superposition along the same line of sight of different clouds at different distances
is not entirely surprising since that line of sight passes only slightly above the
Galactic plane, but certainly warrants caution when considering individual regions
in that direction. For instance, it is quite plausible that some of the patchy structures 
seen in the direction of Serpens on optical plates are in fact associated with foreground  
material in the Aquila Rift.

\section{Conclusions and pespectives}

In this paper, we have presented multi-epoch VLBA observations of EC 95, 
a young stellar object located in the SVS 4 sub-cluster of the Serpens 
molecular core. Our data demonstrate that EC 95 is in fact a tight binary
system with a separation of about 15 mas. The primary (EC 95a) appears 
to be a 4--5 $M_\odot$ intermediate-mass Herbig AeBe protostar, whereas 
the secondary (EC 95b) is most likely a low-mass T Tauri. At radio wavelengths, 
the secondary is on average brighter than the primary. It is, therefore, also likely
to dominate the fairly bright X-ray emission from the system. The primary, 
on the other hand, contributes most of the infrared and of the bolometric 
luminosity of EC 95. This might naturally explain why the extinction to
EC 95 based on infrared observations was systematically found to be 
much larger than the extinction based on X-ray data.

Interestingly, both members of EC 95 appear to be non-thermal radio 
emitters. While a low-mass T Tauri star such as EC 95b is expected
to generate that type of emission, an intermediate-mass such as EC 95a 
is not. We discussed several mechanisms that could explain the presence 
of non-thermal emission on EC 95a, and argue that the observed properties 
of EC 95a might be most readily explained if it possessed a corona powered
by a thin, rotation-driven convective layer.

The trigonometric parallax of EC 95 appears to be $\pi$ = 2.41 $\pm$ 0.02 
mas, corresponding to a distance of 414.9$^{+4.4}_{-4.3}$ pc. We argue that
this implies a distance to the Serpens core of 415 $\pm$ 5 pc, and to the
Serpens molecular cloud of 415 $\pm$ 25 pc. This is significantly larger
than previous distance estimates based on measurements of the extinction
suffered by stars in the direction of Serpens ($d$ $\sim$ 260 pc). A possible 
explanation for this discrepancy is that these measurements correspond to the
distance to clouds associated with the Aquila Rift located in the foreground 
of the Serpens cloud.

Since our observations resolve EC 95 into a binary, future radio monitoring 
of that system will allow us to measure the dynamical mass of that system. 
To our knowledge, this will be the first time that a dynamical mass is obtained for 
a young Herbig AeBe system, and it will enable us to uniquely constrain 
theoretical pre-main sequence evolutionary for intermediate-mass stars. 
The orbital period of the system appears to be 10--20 years, so observations
in the next two decades will be required to obtain an accurate mass estimate.

To Serpens cloud is about 20 pc across on the plane of the sky, so
it is likely to be also about 20 pc deep. Observations similar to those presented
here for a sample of young stars distributed across the cloud would help 
determine the structure of the region, and would allow us to further constrain 
the mean distance to this important region of star-formation.

\end{document}